\documentclass[sigconf]{acmart} 

\AtBeginDocument{
  }

\usepackage{enumitem}
\usepackage{longtable}
\usepackage{booktabs}
\usepackage{tabularx}
\usepackage{xcolor} 
\usepackage{soul}

\AtBeginDocument{
  }

\copyrightyear{2025}
\acmYear{2025}
\setcopyright{acmlicensed}\acmConference[ASSETS '25]{The 27th International ACM SIGACCESS Conference on Computers and Accessibility}{October 26--29, 2025}{Denver, CO, USA}
\acmBooktitle{The 27th International ACM SIGACCESS Conference on Computers and Accessibility (ASSETS '25), October 26--29, 2025, Denver, CO, USA}
\acmDOI{10.1145/3663547.3746385}
\acmISBN{979-8-4007-0676-9/2025/10}

\begin{document}

\title{Understanding How Visually Impaired Players Socialize in Mobile Games}

\author{Zihe Ran}
\affiliation{
  \institution{Communication University of China}
  \city{Beijing}
  \country{China}
}
\email{rougetinge@Gmail.com}

\author{Xiyu Li}
\affiliation{
  \institution{Communication University of China}
  \city{Beijing}
  \country{China}
}
\email{elviaovo@gmail.com}

\author{Qing Xiao}
\authornote{Corresponding Author: Qing Xiao, PhD Student, Human-Computer Interaction Institute, Carnegie Mellon University, Pittsburgh, PA, USA, qingx@andrew.cmu.edu}
\affiliation{%
  \institution{Human-Computer Interaction Institute, Carnegie Mellon University}
  \city{Pittsburgh}
  \state{PA}
  \country{USA}
}
\email{qingx@andrew.cmu.edu}

\author{Yanyun Wang}
\affiliation{
  \institution{College of Media, Communication and Information, University of Colorado Boulder}
  \city{Boulder}
  \state{Colorado}
  \country{USA}
}
\email{mia.wang@colorado.edu}

\author{Franklin Mingzhe Li}
\affiliation{
  \institution{Human-Computer Interaction Institute, Carnegie Mellon University}
  \city{Pittsburgh}
  \state{Pennsylvania}
  \country{USA}
}
\email{mingzhe2@cs.cmu.edu}

\author{Zhicong Lu}
\affiliation{
  \institution{Department of Computer Science, George Mason University}
  \city{Fairfax}
  \state{Virginia}
  \country{USA}
}
\email{zlu6@gmu.edu}

\begin{abstract}
Mobile games are becoming a vital medium for social interaction, offering a platform that transcends geographical boundaries. An increasing number of visually impaired individuals are engaging in mobile gaming to connect, collaborate, compete, and build friendships. In China, visually impaired communities face significant social challenges in offline settings, making mobile games a crucial avenue for socialization. However, the design of mobile games and their mapping to real-world environments significantly shape their social gaming experiences. This study explores how visually impaired players in China navigate socialization and integrate into gaming communities. Through interviews with 30 visually impaired players, we found that while mobile games fulfill many of their social needs, technological barriers and insufficient accessibility features, and internal community divisions present significant challenges to their participation. This research sheds light on their social experiences and offers insights for designing more inclusive and accessible mobile games.
\end{abstract}


\maketitle

\section{Introduction}

Mobile games have experienced a significant rise in popularity over the past five years, becoming one type of the most widely used 
 mobile applications ~\cite{hamid2020driving}. By 2024, the number of mobile game players globally exceeded 3 billion, with revenues reaching \$116.4 billion. The Asia-Pacific region alone contributed 73\% of this revenue, while China and the United States accounted for 49\%, highlighting the gaming preferences of a substantial portion of players ~\cite{su2022few,newzoo2024}.

This growing popularity of mobile games is particularly relevant when considering their potential for enhancing social interactions. The accessibility and convenience of mobile games make them particularly appealing, offering numerous opportunities for socializing. Multiplayer functionality and interactive environments in these games can foster social connections and emotional exchanges ~\cite{zhang2017massively,tong2021players}. This is especially relevant for visually impaired individuals, who face significant challenges in traditional social interactions due to physical and informational barriers ~\cite{thangal2018social,cimarolli2006differences}, especially in Global South without enough physical accessible services. Digital games, including mobile games, have shown promise in reducing isolation and loneliness, enhancing social skills, and promoting psychological well-being ~\cite{granic2014benefits,mittmann2022lina,depping2018designing,bei2024starrescue,grinyer2022massively}.

Despite the positive aspects highlighted in existing HCI research, which often focuses on improving online socialization for visually impaired players ~\cite{ghafoor2024improving,maidenbaum2013increasing,nishchal2023social}, there is a notable gap in examining how social features (ie, within the game world, private chat channels and even develop into online acquaintances or real-life friendships outside of the game) in mobile games specifically impact visually impaired players. While mobile games offer efficient and convenient ways to facilitate social interactions ~\cite{mohd2021apple}, the unique needs and experiences of visually impaired players in this context remain underexplored. Given that many visually impaired individuals in the Global South prefer mobile games ~\cite{maliyo2024,ran2025users}, understanding their socializing experiences in mobile gaming environments is crucial.

Thus, the current study aims to explore how mobile games offer visually impaired players opportunities for socializing and enrich their socializing experiences. We also seek to uncover further possibilities for designing inclusive mobile gaming experiences that are not only entertaining but also facilitate deeper social engagement for visually impaired players. More specifically, we address three key research questions in this study (RQs):
\begin{itemize}
    \item \textbf{RQ1:} What is the socializing experience of visually impaired players when playing mobile games?
    \item \textbf{RQ2:} What challenges do visually impaired players face when socializing in mobile games?
    \item \textbf{RQ3:} What are the expectations of visually impaired players for future mobile game design that supports socializing?
\end{itemize}

To address our research questions, we conducted semi-structured interviews with 30 visually impaired players, all of whom had accumulated over 50 hours of mobile gameplay experience prior to the study. Through this study, we explored the socializing experiences of visually impaired individuals in mobile games, focusing on how mobile games support them in joining communities, both within in-game communities and in building relationships and engaging in social participation beyond the game. (\textbf{RQ1}). Additionally, we identified the social challenges that visually impaired players face due to physiological limitations and how these challenges in gaming socializing reflect the internal divisions and intersecting inequalities faced by visually impaired individuals in the real world (\textbf{RQ2}). Finally, we explored the expectations of visually impaired players, focusing on future design goals for mobile games (\textbf{RQ3}).

This study contributes to Human-Computer Interaction (HCI) and accessibility research in several key ways. First, it repositions mobile gaming as a critical social and cultural domain for visually impaired individuals that extends far beyond leisure and into emotional expression, peer connection, and public participation, especially where access to other social infrastructures is limited. Second, our findings expand the discourse on accessibility by uncovering how social exclusion is not only a result of technical inaccessibility, but also of everyday ableism, identity-based power dynamics, and platform design that implicitly centers normative bodies and behaviors. Third, by foregrounding the lived experiences and aspirations of visually impaired players, we generate actionable insights for designing mobile games that support cross-group interaction, flexible social structures, and pathways to offline inclusion. 

\section{Related Work}
Firstly, we review the existing literature to explore how visually impaired individuals experience gaming (\autoref{2.1}). We then reviewed past research about the barriers faced by visually impaired individuals in socializing and related accessible tools to overcome such barriers (\autoref{2.2}). Next, we examined studies about game socializing, with a particular focus on mobile games (\autoref{2.3}).

\subsection{Gaming and Individuals with Visual Impairments}\label{2.1}

Previous research has found that games specifically designed for visually impaired players are often simplified and lack the enjoyment and complexity of mainstream games \cite{ran2025users,andrade2019playing,gonccalves2020playing,smith2018rad}. These games are generally simplified versions of those enjoyed by sighted players \cite{allman2009rock,kim2011tapbeats,miller2007finger,morelli2010vi}, requiring players to perform basic actions based on commands. For instance, Allman et al. reported on audio games modified for visually impaired players that rely on tactile and auditory feedback for gameplay \cite{allman2009rock}.

Some game developers and researchers have acknowledged the importance of making digital entertainment more inclusive and have developed accessible games or assistive tools for visually impaired players \cite{yuan2008blind,sepchat2006semi}. For example, Nair et al. created Surveyor, an exploration tool that helps visually impaired players enhance their ability to discover objects within a virtual map, improving their gaming experience ~\cite{nair2024surveyor,nair2021navstick}. Metatla et al. designed an inclusive game robot in collaboration with visually impaired children to help them develop certain skills while playing ~\cite{metatla2020robots}. Despite these efforts, visually impaired players have not fully integrated into mainstream gaming. Andrade et al. highlighted that visually impaired players face concerns about self-expression and being left behind by emerging technologies ~\cite{andrade2019playing}. Furthermore, many visually impaired players express a strong desire to play mainstream games \cite{bar2018tangicraft,gonccalves2020playing}. For example, Gonçalves et al. studied how visually impaired players use spatial cues and sound to play mainstream games \cite{gonccalves2020playing}. 

Despite these efforts, visually impaired players continue to face challenges due to a lack of information, requiring greater cognitive effort to play games alone without social support, which results in less enjoyable experiences. This need also reflects a psychological desire among visually impaired players to connect with mainstream society. Indeed, prototypes that allow blind and sighted users to play games fairly through dual-mode interfaces have been developed \cite{plimmer2008multimodal,winberg2004assembling}, but their widespread adoption and applicability still have room for development. Further design exploration is needed to expand social possibilities for both visually impaired and sighted players in mobile games.

\subsection{Socializing Among Visually Impaired Individuals}\label{2.2}

In this section, we reviewed previous studies on the socializing barriers of visually impaired individuals, including psychological barriers, information barriers. We also reviewed scholars' attempts to improve accessibility in socializing for visually impaired individuals.

\subsubsection{Psychological Barriers to Socializing for Individuals with Visual Impairments}

Numerous studies have explored the psychological barriers visually impaired individuals face in socializing contexts ~\cite{perez2021consequences,cha2024you,oh2016effects}. In professional and organizational contexts, past research highlights the subtle dilemmas that visually impaired professionals face such as lacking accessible tools, additional labor for image management and workplace discrimination \cite{cha2024you}. For example, Cha et al. found that conference activities and the software tools used during these events are often inaccessible, requiring individuals with visual impairments to put in additional effort. This made them frequently strive to conceal their disabilities during conferences, managing how visible their impairments are to others \cite{cha2024you}. 

Visually impaired individuals also constantly employ various strategies to reshape their self-identity and social image ~\cite{li2022feels}. Specifically, Li et al. discovered that visually impaired individuals are similarly influenced by societal norms regarding makeup. Despite facing numerous challenges, such as a lack of learning materials and difficulties in distinguishing makeup products, they actively engage in makeup to influence their socializing image ~\cite{li2022feels}. In many occasions, disability remains largely stigmatized, with many viewing visually impaired individuals as a group in need of fixing or pity, which further reinforces psychological barriers for visually impaired individuals in interpersonal communication with other \cite{reynolds2017d,li2021choose,grue2016social}. 

Previous research has examined how visually impaired people interact with others in different social environments \cite{smedema2010relationship,oh2016effects,bennett2018teens,kianpisheh2019face,bandukda2024context}. For example, Smedema et al. found that more online social interaction was linked to slightly better overall well-being among visually impaired individuals \cite{smedema2010relationship}. However, Oh et al. pointed out that many accessibility tools do not work well with newer devices, which can make communication more difficult and cause anxiety or other forms of psychological stress \cite{oh2016effects}. These studies suggest that it is not just personal motivation that affects the social lives of visually impaired people, while the quality and availability of accessibility technology also play a major role in shaping their ability to connect with others.

In this context, social digital platforms, especially multiplayer mobile games, deserve closer attention, as they combine leisure, communication, and participation in particularly immersive ways. Building on these insights, our research focuses on the social experience with challenges faced by visually impaired individuals in mobile gaming environments and investigates how game design can better support inclusive, meaningful interactions.

\subsubsection{Information Barriers in Online Social Interactions for Visually Impaired Individuals}
Recent studies have drawn attention to the information barriers that visually impaired individuals frequently encounter on social media platforms \cite{wang2021revamp,jiang2024s,segal2024socialcueswitch,ning2024spica,bandukda2024context}. For instance, Sharevski and Zeidieh found that visually impaired users often struggle to detect misinformation warnings, which directly impacts their ability to make informed decisions online \cite{sharevski2024assessing}. Other research challenges common assumptions about visual content: Bennett et al. showed that visually impaired individuals remain interested in visual media and wish to participate in visual forms of social interaction \cite{bennett2018teens}. Similarly, Voykinska et al. revealed how users employ creative strategies to interpret or work around inaccessible visual content on social networking platforms \cite{voykinska2016blind}, demonstrating both resilience and adaptability. These findings highlight a key tension: While visually impaired individuals actively seek social participation and engagement, they continue to face information barriers that limit their access to information and reduce the quality of their online socializing experience.

To address these persistent information barriers, although not focused on social experience, HCI researchers have proposed a range of technological solutions aimed at improving information accessibility for visually impaired users. For example, Wu et al. introduced the use of automatic alternative text (AAT), an AI-based tool that generates image descriptions to help visually impaired individuals better understand visual content \cite{wu2017automatic}. Although such tools represent meaningful progress, many users still face significant challenges in accessing critical information and fully participating in online social environments \cite{wu2017automatic,bennett2018teens,sharevski2024assessing}. These barriers not only restrict access to knowledge and digital services, but also hinder emotional expression, interpersonal connection, and a broader sense of social belonging.

Despite these growing insights, much of the existing research is based on traditional social media platforms, and less is known about how visually impaired individuals experience newer forms of online interaction, such as mobile gaming, that combine entertainment with real-time communication. To address this gap, our study investigates how information barriers and design choices in mobile games shape the social and emotional experiences of visually impaired players.

\subsubsection{Accessibility Efforts to Improve Social Experiences}
With the advancement of technology, there has been growing research interest in supporting the social participation of visually impaired individuals in digital environments \cite{segal2024socialcueswitch,ji2022vrbubble,collins2023guide}. For instance, Maidenbaum et al. developed "EyeCane," a virtual mobility aid designed to enhance the accessibility of social activities in virtual spaces \cite{maidenbaum2013increasing}. Collins et al. further highlighted the unique social preferences of visually impaired users in such environments, emphasizing the important role of sighted assistants in enabling meaningful interactions \cite{collins2023guide}. Other efforts, such as the work by Komal et al., explored conversational accessible technologies to support real-time communication for visually impaired individuals \cite{ghafoor2024improving}. Similarly, Nishchala et al. proposed a machine-learning-enhanced social networking pipeline to help visually impaired users navigate and engage more effectively on mainstream platforms \cite{nishchal2023social}.

These studies underscore the increasing attention to accessibility in online social interactions. However, most of this work focuses on general social settings, such as messaging platforms or virtual communities, rather than on specific, immersive social contexts like mobile gaming. Given that mobile games often blend real-time communication, play, and collaboration, they present distinct challenges and opportunities for accessibility. Our study builds on this emerging body of work by examining how the design of mobile games shapes the social and emotional experiences of visually impaired individuals, and by identifying barriers and possibilities unique to this increasingly popular form of digital interaction.

\subsection{Socializing in Gaming}\label{2.3}

In human-computer interaction and accessibility studies related to online games, the social functions of games, such as improving social skills and fostering emotional interactions, have received considerable attention ~\cite{bei2024starrescue,depping2018designing,mittmann2022lina,grinyer2022massively,audiffred_hinojosa2022post}, as has the design of games with strong social attributes ~\cite{rosenqvist2018meteorquest,robinson2020designing}. 

\subsubsection{The Psychological Significance of Social Gaming}

A growing body of research has highlighted the positive psychological effects of social interactions in digital games, particularly in supporting mental health and emotional well-being \cite{depping2018designing,wu2023mr}. For instance, Bei et al. developed StarRescue, a cooperative turn-based tablet game designed to enhance social communication skills, such as prompting, negotiating, and task distribution, among children with autism \cite{bei2024starrescue}. Depping et al. explored how in-game interactions fulfill basic psychological needs like relatedness and help alleviate feelings of loneliness \cite{depping2018designing}. The COVID-19 pandemic further intensified academic interest in the emotional and social potential of games, with studies showing how massively multiplayer online role-playing games (MMORPGs) helped mitigate social isolation and support psychological resilience during periods of physical distancing \cite{grinyer2022massively,audiffred_hinojosa2022post,yuan2021tabletop}.

At the same time, scholars have also begun to critically examine the ambivalent and sometimes exclusionary social dynamics of digital games ~\cite{deng2024they}. While games can foster inclusion and belonging, they can also reproduce social hierarchies, foster in-group favoritism, and exclude those who do not conform to dominant norms of communication or gameplay \cite{tushya2023social,ducheneaut2006alone}. Prior studies have noted how tightly-knit in-game cliques may discourage interaction with outsiders ~\cite{ward2022network}, especially when players rely on pre-existing friendships, fast-paced coordination, or visual cues to form teams \cite{ran2025users}. These dynamics can be potentially challenging for players with disabilities, who may be perceived as "less competent" due to differences in interaction speed or interface use, leading to subtle forms of social marginalization and self-exclusion \cite{ran2025users}. These insights motivate our investigation of how visually impaired individuals navigate both the opportunities and obstacles embedded in mobile game–based social interaction.

\subsubsection{Games with Strong Social Attributes}
Building on Gonçalves et al.’s research on digital games, social games can be defined through four dimensions: (1) Social as non-solitary—games in which more than one player actively participates and where gameplay is designed to foster interaction; (2) Social in the interactions during and outside the game—games that reflect sociality both through in-game character interactions and through the social relationships among players beyond the game context; (3) Social in the outcomes—games that produce social effects or influence are considered “social”; and (4) Social inherent to gaming—games that offer built-in social functionalities ~\cite{gonccalves2023social}. 

Based on this framework, we define the scope of social games in our study as games that enable real-time or asynchronous social interaction (e.g., chatting, collaboration, gifting) and encourage the development of social relationships beyond the gameplay itself. Within this domain, HCI researchers have explored how to amplify games’ social potential through novel interaction techniques and emotionally resonant design. For example, Hirsch et al. proposed using asynchronous heartbeats to enhance a sense of social co-presence in single-player VR games \cite{hirsch2023my}. Rosenqvist et al. introduced MeteorQuest, a mobile location-based game that encouraged shared exploration among family members \cite{rosenqvist2018meteorquest}. Robinson et al. developed a two-player endless running game requiring synchronized input, which fostered social intimacy through collaborative gameplay \cite{robinson2020designing}.

These studies about social game illustrate the breadth of innovation in social game design, ranging from emotional storytelling to embodied cooperation. However, most of this work has implicitly assumed sighted players as the default user group. For visually impaired individuals, many of these social mechanics, particularly those relying on visual cues, spatial orientation, or fast-paced coordination, pose unique accessibility challenges. Despite the increasing attention to social interaction in games, limited research has examined how these experiences are shaped by the intersection of disability and game design. 

\subsubsection{Socializing Experience in Mobile Gaming}
Previous research has shown that the social attributes of mobile games are deeply intertwined with players' overall experiences and long-term engagement \cite{zhang2020mobile,vuijk2021patrec}. Paul et al. identified social interaction, such as cooperation, competition, and community belonging, as a key motivational factor driving continued participation in mobile games \cite{paul2008socializing}. These interactions are not merely supplementary features but are often central to how players derive meaning, enjoyment, and connection from gameplay.

A growing body of literature has explored how social design elements shape emotional and behavioral outcomes in mobile gaming. For instance, studies on location-based games (LBGs), such as Pokémon GO, have demonstrated how mobile games can bridge online and offline interactions, encouraging co-presence, collaboration, and spontaneous encounters in physical spaces \cite{saaty2022pokemon,wu2023mr}. Saaty et al. emphasized the importance of reimagining these social experiences in the context of remote and asynchronous gaming, particularly in a post-pandemic era where physical proximity is no longer guaranteed \cite{saaty2022pokemon}.

Beyond LBGs, other forms of mobile games, such as multiplayer real-time strategy games, social simulation games, and rhythm-based games, have embedded various social mechanisms like team-based matchmaking, in-game gifting, and asynchronous messaging, all of which aim to foster social presence and emotional exchange. These features contribute to players’ sense of inclusion, shared purpose, and relational continuity across gaming sessions \cite{zhang2020mobile}.

While these studies reflect a growing academic recognition of the social dimensions of mobile gaming, they tend to focus on sighted players and normative assumptions about social behavior, accessibility, and interface interaction. Our study extends this line of inquiry by examining how visually impaired individuals experience, navigate, and reinterpret the social mechanics embedded in mobile games. We explore both the affordances and limitations of current game designs, with attention to how social interactions are shaped not only by in-game features but also by broader accessibility infrastructures. In doing so, we seek to expand the conversation around mobile game sociality by centering players whose experiences have historically been marginalized in both game design and research.

\section{Method: Semi-structured Interviews}

Given the research gap in the social experiences of visually impaired players in mobile games, we aims to enhance our understanding of how visually impaired players engage in mobile game socialization. 

\subsection{Participants: Individuals with Visual Impairments (N=30)}
To better understand the social experiences, challenges, and expectations of mobile game players regarding the design of future mobile games, we conducted semi-structured interviews with 30 visually impaired participants. All visually impaired participants had spent over 50 hours playing the games prior to this study and had used social features related to the games. In our study, we did not track participants' gameplay time directly by ourselves; instead, we relied on their self-reported estimates. This approach was chosen to respect participants' privacy, while still providing a meaningful indicator of their familiarity with the game. This study received approval from the Institutional Review Board (IRB) at the affiliated university.

Visually impaired players were recruited through forums and communities specifically for visually impaired players. The recruitment criteria for visually impaired players were: 1) being at least 18 years old, 2) having a visual impairment, and 3) having more than 50 hours of gaming experience. Detailed demographic information is presented in Table~\ref{tab:participants of visually impaired}.

\begin{table*}[h]
 \caption{Information of visually impaired participants. }
  \label{tab:participants of visually impaired}
\setlength{\tabcolsep}{3pt} 
\scalebox{0.8}{
\begin{tabular}
{>{\centering\arraybackslash}p{0.7cm} 
                >{\centering\arraybackslash}p{1.6cm} 
                >{\centering\arraybackslash}p{1.2cm} 
                >{\centering\arraybackslash}p{2.5cm} 
                >{\centering\arraybackslash}p{3cm} 
                >{\centering\arraybackslash}p{2.8cm} 
}
    \hline
    \textbf{ID} & \textbf{Age Range} & \textbf{Gender} & \textbf{Vision Conditions} & \textbf{Weekly Playtime} & \textbf{Number of Mobile Games} \\
    \hline
    V1 & 20-30 & Male  & Blind & 20+ hours & 6-10 \\
    V2 & 20-30 & Male  & Blind & 10+ hours & 1-5 \\
    V3 & 20-30 & Male  & Blind & 15+ hours & 10+ \\
    V4 & 20-30 & Female & Low Vision & 20+ hours & 10+ \\
    V5 & 18-20 & Female & Blind & 20+ hours & 6-10 \\
    V6 & 20-30 & Male & Low Vision & 5+ hours & 1-5 \\
    V7 & 20-30 & Male & Blind & 25+ hours & 6-10 \\
    V8 & 40-50 & Male & Blind & 10+ hours & 6-10 \\
    V9 & 18-20 & Male  & Blind & 10+ hours & 1-5 \\
    V10 & 20-30 & Male  & Low Vision & 5+ hours & 1-5 \\
    V11 & 20-30 & Male  & Blind & 25+ hours & 10+ \\
    V12 & 18-20 & Male  & Low Vision & 20+ hours & 6-10 \\
    V13 & 30-40 & Female  & Blind & 10+ hours & 6-10 \\
    V14 & 30-40 & Male  & Blind & 20+ hours & 10+ \\
    V15 & 20-30 & Female & Blind & 15+ hours & 6-10 \\
    V16 & 20-30 & Male  & Blind & 15+ hours & 10+ \\
    V17 & 20-30 & Female & Blind & 35+ hours & 10+ \\
    V18 & 18-20 & Male & Blind & 10+ hours & 6-10 \\
    V19 & 20-30 & Male & Blind & 25+ hours & 10+ \\
    V20 & 30-40 & Male & Low Vision & 5+ hours & 1-5 \\
    V21 & 20-30 & Male  & Blind & 5+ hours & 1-5 \\
    V22 & 20-30 & Male  & Blind & 10+ hours & 1-5 \\
    V23 & 30-40 & Male  & Blind & 35+ hours & 10+ \\
    V24 & 20-30 & Male & Blind & 15+ hours & 5-10 \\
    V25 & 20-30 & Male & Blind & 20+ hours & 10+ \\
    V26 & 18-20 & Female & Blind & 10+ hours & 6-10 \\
    V27 & 20-30 & Male & Low Vision & 5+ hours & 6-10 \\
    V28 & 20-30 & Male  & Blind & 10+ hours & 10+ \\
    V29 & 20-30 & Male & Blind & 5+ hours & 1-5 \\
    V30 & 20-30 & Male & Blind & 15+ hours & 10+ \\ 
    \hline
\end{tabular}
}
\end{table*}

\subsection{Study Procedure}
We conducted one-on-one semi-structured interviews with 30 participants, each lasting approximately two hours. We first collected basic demographic information from participants, including age, gender, visual condition, and occupation. 

For the visually impaired players, we then asked about their socializing needs in daily life and whether the socializing systems in mobile games met or hindered these needs. For instance, we explored whether they made friends through mobile games, what types of social relationships they developed in mobile games, and their attitudes toward in-game communication. We also asked how their visual condition influenced their social preferences and how these preferences manifested within gaming contexts. Next, we asked visually impaired players to describe both positive and negative socializing experiences in mobile games, with an emphasis on how their visual condition played a role in these interactions. For instance, we asked visually impaired participants whether they felt the need to disclose their disability during social interactions, whether they experienced unfair treatment, and how communication occurred between sighted and visually impaired players in mobile games. Finally, we explored their expectations for the future of socializing systems in mobile games, including the ethical considerations and design goals of specific features they believe should guide the design of social features in future mobile games.

\subsection{Data Analysis}

All interviews were conducted in Mandarin Chinese by the first and second authors. Each interview was audio-recorded and transcribed verbatim. For analysis, the transcripts were translated into English. We were mindful of the risks of translation bias, particularly the potential loss of cultural nuance, emotional tone, or context-specific meanings during the translation process. To mitigate this, the translation was carried out collaboratively by bilingual researchers familiar with both linguistic conventions and the cultural context of the participants. Where ambiguity arose, we revisited the original Chinese transcripts to ensure accurate interpretation. In several cases, key phrases were discussed among the research team to preserve participants’ intended meanings and to avoid over-simplification or culturally inappropriate renderings.

We employed thematic analysis \cite{braun2019reflecting}, chosen for its flexibility and capacity to surface both explicit and latent patterns in qualitative data. Throughout the research process, we drew on prior HCI approaches to reflexive thematic analysis (e.g., \cite{xiao2025let,xiao2025might,kuo2023understanding,fan2025user}), engaging in reflexive practices to critically examine our own positionality and its influence on the data. As sighted researchers studying the experiences of visually impaired participants, we recognized the risk of unintentionally reinforcing ableist assumptions or framing impairment through a deficit lens. To counter this, we avoided predefining accessibility solely as a technical challenge and instead centered participants’ own articulations of inclusion, agency, and play. During coding and theme generation, we paid particular attention to preserving the diversity of participants' perspectives, avoiding generalizations that flatten the heterogeneity of visually impaired players’ experiences.

Three researchers analyzed data from 30 interviews, totaling approximately 78 hours of recordings. After iterative coding, the team collaboratively reviewed the transcripts to develop a shared understanding of emerging patterns, generating 912 codes. We then employed affinity diagramming \cite{pernice2018affinity} to organize codes into clusters of related concepts and to identify and conceptualize key themes. These themes encapsulate the gaming experiences of visually impaired players, the role of gaming in shaping socializing, and the considerations for designing future socializing systems in mobile games. 

\subsection{Positionality Statements}

Our research team brings together a range of lived experiences and disciplinary perspectives. While researchers in our study are sighted, two authors have engaged in long-term collaboration with blind and low-vision communities and volunteered in related non-profit organizations, including participating in accessibility initiatives and co-playing digital games alongside community members. One co-author is an HCI researcher whose work centers on blind and low-vision individuals. Their presence helped counteract the risk of sighted-normative assumptions and supported a more situated, empathetic approach to the data. Another member of the team brings prior ethnographic research experience. With training in reflexive ethnography, this insights helped guide the team’s ongoing reflexivity, particularly in interpreting ambiguity, improvisation, or silence in participant narratives. Additionally, two researchers have longstanding experience in game research and design. Their expertise provided crucial insight into the dynamics of multiplayer systems, player collaboration, and game mechanics, allowing the team to interpret participants' gaming experiences in relation to broader design paradigms. Throughout the research process, we engaged in regular reflexive discussions to examine how our own positions, experiences, and assumptions shaped the questions we asked, the meanings we constructed, and the ways we represented participant voices.

\section{Findings: Socializing in Mobile Games for Visually Impaired Players (RQ1)}

Previous studies have explored the social experiences of visually impaired individuals. In this section, we will introduce the social experiences of visually impaired players in mobile games from four aspects: information network(\autoref{4.1}), game communities (\autoref{4.2}), emotional connection (\autoref{4.3}), and social skills development (\autoref{4.4}).

\subsection{Mobile Games as Informal Information Networks for Visually Impaired Players}\label{4.1}

Based on the accounts of our visually impaired participants (N=30), many described significant delays in accessing timely information, particularly around pop culture and social trends. As V4 noted, \textit{"Although I'm in my twenties, I feel like an older person who knows nothing about the popular culture that today's youth enjoy."} This difficulty stems not only from the lack of screen-reader–compatible content, but also from the visual-first design of mainstream information platforms in China, such as Weibo, Douyin, and WeChat. These platforms often present information through video thumbnails, images with embedded text, or scrolling banners, where these formats that are inherently inaccessible to screen-reading technologies. Furthermore, real-time popular discourse is frequently mediated through ephemeral visual memes, livestream interactions, or short-form video comment threads, which are difficult to capture or interpret without direct visual access. This information lag not only limits access to shared cultural knowledge, but also creates barriers to social participation. Participants expressed frustration at being left out of peer conversations, unable to keep up with real-time trends, and feeling socially disconnected as a result because they lack a shared understanding of popular culture with sighted peers.

Participants emphasized that with limited access to timely content, they turn to mobile game communities not just for play, but as a crucial space for information exchange and social inclusion. For V19, the social interactions in games have become more valuable than the games themselves: \textit{"Every day at home, I have nothing to do, and I cannot access new information. But in the game community, I can hear various stories from other players. I long to hear others' stories. It makes me feel connected to the world, rather than isolated and trapped, unable to go outside."} 

The lack of accessible information infrastructures leaves many visually impaired individuals feeling disconnected from public life. As V14 shared: \textit{"Sometimes, I feel like I'm an isolated island. I have no job, no school to attend. I stay home every day. When my family goes out to work, I often remain in silence."}

These narratives underscore that for many participants, the effort to access information is not just about content but also about maintaining social presence, keeping up with peers, and resisting isolation. Mobile game communities thus become informal lifelines where information, emotion, and identity circulate in accessible ways.

However, social interactions within mobile games have somewhat mitigated this issue. In mobile games, visually impaired players can make friends and chat with one another, allowing for a natural exchange of information. Through these interactions, they can keep each other updated on the things they know. V14 added the help of mobile game socializing, saying, \textit{"Before, when I was home alone with no one to talk to, I knew nothing. But when I started playing mobile games and listened to other visually impaired individuals chatting, I learned about their life experiences. That brought me a lot of comfort."}

Among our visually impaired participants, it is noteworthy that 22 out of 30 considered the information access gained through social interactions as one of the most valuable aspects of socializing in mobile games, even surpassing the experience of making friends (n=19) and emotional connections (N=19). This highlights the information barriers they face in real life and underscores the critical role of mobile games as an alternative means to bridge this gap. V18 illustrated this point, saying, \textit{"For me, making friends isn't important; I've grown accustomed to loneliness. But what really worries me is being isolated, losing connection with society, and being completely unaware of the outside world. I’m glad to be part of the mobile gaming community for socializing, as it has allowed me to learn many new things."}

\subsection{Game Communities: Guild, in-Game Alliance and Voice Chat Rooms}\label{4.2}

In addition to world channels, mentioned by our visually impaired participants, many mobile games feature more private social systems such as guilds, factions, or alliances. These smaller groups often have stronger mutual interests, more frequent information exchanges, and closer relationships. Within these groups, players often develop different types of relationships, such as mentor-mentee or sibling-like bonds in the game, frequently communicating through public guild channels. V18 sought game strategies through guilds. V18 thoroughly enjoys engaging in discussions about the game with other members of the guild. He finds great satisfaction in learning from others' experiences and strategies while also sharing his own insights and knowledge. This  mutual support makes him feel as though he is a part of a community where everyone shares a common interest and passion for the game. It fosters a sense of belonging and camaraderie, as they collectively strive to improve, learn, and progress together. 

V18 said, \textit{"When we are stuck on a difficult challenge and work together to find a solution, I feel like I am part of a team. We are striving toward a common goal. In my real-world life, I have rarely had this kind of experience—where someone wants to be on a team with me. But in mobile games, I have found the team where I belong."} V3, who once led an alliance in a mobile game designed for the visual impairment, managed 13 group chats within the game. The internal chat systems allowed V13 to stay fully informed about conflicts between guilds or the latest game updates. V13 proudly said, \textit{"Can you imagine? I manage 13 different groups, and each one is filled with friendly people who really enjoy chatting within the group. I can also keep track of the latest guild updates, notify everyone, and share game strategies. We fight like brothers and sisters in the guild."}

Beyond in-game channels and alliances, many visually impaired players actively participate in external game-related communities, which offer additional platforms for information exchange and support. 18 visually impaired participants in the interviews were part of mobile game-related social chat groups, with nine participants actively engaging in discussions to share information. 

For visually impaired participants who actively share information and engage in chats, the experience is liberating: \textit{"I feel like I've found a place where I can speak freely, where no one cares about my disability" }(V3). \textit{"I enjoy chatting in the group. No one discriminates against me; instead, I feel that my words are valued" }(V7).

For those who prefer to quietly observe others' conversations without initiating messages, these group chats still provide substantial support. \textit{"Seeing everyone chat in the group makes me feel like I am also a part of them" }(V15). This more reserved way of joining a mobile game community is a common approach among visually impaired individuals.

Among the 17 visually impaired people we interviewed, many reported that they frequently browse community messages passively without actively posting. This behavior reflects their social habits in the real world: \textit{"We mainly listen to sighted people speak rather than express our own opinions. Whenever I voice my thoughts, I am always contradicted"} (V8). \textit{"Sometimes sighted people don't include me in their conversations, and I feel excluded"} (V11). 

These reflections suggest that passive engagement is not merely a matter of individual preference but is shaped by broader patterns of social marginalization. Even so, participants emphasized that simply observing others’ conversations in chat groups still fostered a sense of connection. Within this context, passive observation becomes a protective and empowering strategy, offering a kind of social "buffer zone" where individuals can gauge the tone of the group, understand unspoken norms, and gradually build confidence. As V15 put it,\textit{ “Seeing everyone chat in the group makes me feel like I am also a part of them.” }This illustrates that both active and passive modes of participation offer meaningful pathways to belonging and social inclusion in mobile gaming communities.

\subsection{Emotional Connections in Mobile Gaming}\label{4.3}

For the majority of visually impaired participants (n=19), emotional connection was a primary motivation for social interaction. Many visually impaired participants (n=24) expressed difficulty in joining conversations with the sighted people. As noted by V12, a high school student: \textit{"A lot of popular jokes are based on visual information, Therefore, I actually don’t understand the jokes that sighted people tell. Sometimes, like at home, everyone is laughing out loud when talking about something."} V12 further explained, \textit{"I don’t understand why it’s so funny. I can only laugh along so as not to seem out of place. But at that moment, when everyone was laughing, I actually wanted to cry. I hated myself for not being able to keep up with these topics and not understanding why everyone was laughing. I found that I couldn’t integrate into everyone’s emotions. Even if they regarded me as a friend and family. I still couldn’t empathize with them because I didn’t understand these jokes."} V12 concluded, \textit{"Maybe it’s difficult for sighted people to understand this, and they may think that all problems can be solved by just being friends. That’s not the case. Socializing itself means empathy. If we can’t do this, if we can’t understand each other’s emotions, we will always be people from two different worlds."}

V6, V29, and V30, expressed a desire for conversation topics that transcend visual boundaries for the sighted, fostering deeper emotional connections. Mobile games have become a great medium for findings alternative conversation topics no matter for visually impaired people or sighted people, while V30 shared a particularly vivid experience: \textit{"I recently played a popular new game and shared my thoughts on the game design. A lot of people, both sighted and visually impaired, came to discuss with me, and we had a great time."} 

For many visually impaired individuals, online companionship within games has become an essential part of their daily lives, providing emotional comfort and a sense of social belonging. Participants (n=10) mentioned how online games helped overcome the difficulties of organizing in-person gatherings, which can be challenging for this community. For example, V5, V22, and V26 engaged in voice chat during gaming sessions and offer senses of companionship for each others. 

V5 often feels isolated due to limited contact with the outside world while mobile game socializing reduce such feelings of isolation: \textit{ "I enjoy playing games with large groups of people, like MMO games. When we raid together, we use voice chat to talk during the battle. After the raid, if no one leaves the group, we often chat about other things, whether it's about the game or how we felt about the last raid. Chatting with everyone makes me feel less lonely, and I love the warm, happy feeling it gives me." }

Given the accessibility challenges of offline socializing for our sampled participants, mobile games have become a crucial platform for visually impaired individuals to strengthen interpersonal connections and meaningful interpersonal emotions that contribute to their sense of belonging and well-being. V17's experience serves as a typical example: during a period of unemployment, he made like-minded friends through gaming, \textit{"My in-game friends are important emotional support for me, and talking with them makes me feel accepted by society."} Additionally, some visually impaired individuals develop romantic relationships through gaming. Some enter games with the hope of finding a partner, while others meet and form relationships through gaming interactions. The variety of gameplay options in mobile games socializing offers methods on overcoming the challenges of real-life relationships building. V10 discovered a game dubbed the "Cupid game" after seeing players send love letters to each other on a game-related public account, and he downloaded the game hoping to find romance. V15, on the other hand, incorporated in-game love stories into his self-created novel after meeting his girlfriend through the game’s matchmaking system. Their mutual cooperation and long-term communication eventually blossomed into a romantic relationship.

\subsection{Social Skill Development and Social Inclusion in Mobile Game Socializing}\label{4.4}

In China, visually impaired people are usually educated and trained in a specialized education system, where they only have access to other people with disabilities on most occasions, especially visually impaired people. This makes it difficult for them to come into contact with their sighted peers. This often leaves them with a lack of understanding of socialization strategies in mainstream Chinese society, which is predominantly full of sighted people and inaccessible. V14, who works as a massage therapist, experienced this gap when he began socializing with sighted individuals in his first year on the job: \textit{ "The communication norms of visually impaired people are different from those of sighted people. Visually impaired individuals tend to express things directly, but sighted people use both verbal and non-verbal cues, like body language, to convey their thoughts. Their social etiquette, such as handshakes, bowing, and exchanging business cards, is unfamiliar and difficult for us to learn."}

V21 and V25, who also noticed these differences, turned to role-playing mobile games to bridge this gap. In these games, players simulate real-world social scenarios, taking on various roles and learning how to apply different etiquette rules in contexts such as polite greetings in formal settings or appropriate communication during negotiations. V21 stated: \textit{"By learning, imitating, and practicing in mobile games, even that it is just the virtual socializing in games, I hope to become more confident and composed in future socializing."}

Mobile games, as highly simulated environments of real-world relationships, offer all players a rich and safe online space to observe, learn, and practice social strategies. Team-based missions, free trade markets, and chat rooms mirror real-life social environments, allowing players to learn how to communicate, negotiate, or resolve conflicts. V22, a dedicated mobile player who spends 7 to 8 hours a day immersed in gaming, shared his experience: \textit{"Before playing mobile games, I didn't trust people easily. But in mobile games, I had to fight alongside teammates, and I gradually learned how to build trust with others. I’ve also encountered conflicts, like when I took the victory spoils without consulting my teammates. This led to disagreements, but in the end, we understood each other and gave constructive feedback. I learned to be more empathetic." }

This process of building trust and learning collaboration through gaming exemplifies how social skills can develop. Games not only offer a safe virtual space but also provide visually impaired players the opportunity to overcome real-world social barriers. As their social abilities improve in-game, they become more confident in navigating real-life social interactions. 

As games grow in scale and resources, they can extend to offline activities such as visually impaired eSports competitions, which has already been attempted by a gaming company in China in October 2021. These activities serve as platforms for visually impaired individuals to showcase their abilities, enhance their sense of efficacy, and increase their participation in society. V19, who participated in a visually impaired eSports competition, noted: \textit{"The offline eSports experience is completely different from playing online—it’s more intense and immersive. After the competition, we had dinner together, exchanged contact information, and it felt like being part of a big family. It's expanded my social circle, in fact, I never thought my social skills would progress so quickly in games and in actual interactions in reality"}

The rise of visually impaired eSports competitions and other offline activities demonstrates that mobile games also provide visually impaired players with more opportunities to showcase their abilities and integrate into society. These events not only boost self-efficacy but also strengthen social participation and belonging. We found that social functions of games help visually impaired players overcome real-world barriers while building a broader social network and support system.

\section{Findings: Socializing Challenges within Mobile Games (RQ2)}

In this section, we will discuss the challenges visually impaired players face when engaging with in-game socializing. These challenges include technical limitations (\autoref{4.2.1}), and how real-world social divisions manifest within mobile gaming environments (\autoref{4.2.2}).

\subsection{Technical Limitations Leading to Inability to Socialize with Sighted Players}\label{4.2.1}

Currently, games designed specifically for visually impaired players are often fully accessible. However, these games are generally shared by the visually impaired players, while mainstream mobile games in China typically lack accessibility features such as screen reader (TTS) compatibility, audio cues, and location identification, naturally segregating Chinese visually impaired players from sighted players. V25 expresses disappointment with this divide: \textit{ "Mobile games designed for the visually impaired only attract these visually impaired players, creating a closed social circle. The more interesting mainstream games lack accessibility features, and even when visually impaired friendly versions are made, the social systems don’t integrate. As a result, it’s difficult for visually impaired and sighted players to game together, let alone communicate."}

V27, one of the few visually impaired players who explores mainstream games without accessible tools, notes the challenges: \textit{"These mainstream mobile games without any tools require significant time and energy to navigate, for example, you need to constantly run around the map and collide with surrounding obstacles to determine the general layout and boundaries of the map, and you usually end up performing poorly. The only social experience you have is getting berated by teammates."}

Even when able to communicate with sighted players, visually impaired players worry about how accurately they can express themselves. While speech recognition technology is commonly used in mobile games, and visually impaired users are comfortable with converting speech to text, the process isn’t without issues. Some interviewees (n=3) mentioned they often replay their voice-generated text multiple times to check for errors or homophones. V4 expresses concern: \textit{ "We blind people rely on text-to-speech (TTS) systems, which works well for us. But sighted people read the text, and typos look awkward or confusing to them. It might even give the impression that we are uneducated."}

V6 echoes this frustration, explaining that they double- or triple-check their text to ensure its accuracy but still find it impossible to avoid mistakes entirely. This, they admit, is a source of ongoing frustration.

\subsection{The Intersection of Real-World Social Dilemmas and In-Game Socializing}\label{4.2.2}
In this section, we explore the reflection of real-world social challenges within the socializing of mobile games, specifically manifesting as internal divisions within the visually impaired community due to differences in vision and information access (\autoref{5.3.1}), as well as intersectional inequalities resulting from social identities and power structures within mobile gaming environments (\autoref{5.3.2}).

\subsubsection{Internal Segmentation within the Visually Impaired Community in Mobile Gaming}\label{5.3.1}

While mobile games offer opportunities for social connection, our findings reveal complex dynamics of internal segmentation within the visually impaired gaming community. These divisions are shaped not only by shared impairments but by differences in the degree, onset, and embodied experience of visual impairment, which influence how individuals navigate game-based social spaces and access information within them.

Physiological differences, such as being congenitally blind, becoming blind later in life, or having low vision, constitute a primary layer of stratification during mobile game socializing. Participants often formed tighter bonds with those who shared similar sensory histories, as this commonality shaped not just communication preferences but entire modes of perception and world understanding. V20, who is low vision, described how residual visual cues gave her greater confidence in open-ended conversations with strangers: \textit{"In games that emphasize social interaction, I often approach players to chat about any topic. My slight vision impairment gives me more confidence in social situations [compared to blind players]."} 

In contrast, V15, who is congenitally blind, expressed a sense of detachment from players with different visual histories: \textit{"I rarely chat with those who became blind later or have low vision. They have seen the world and often include visual information, such as colors, in their conversations, which I cannot imagine. This makes it difficult to relate to them and find common topics."} 

This reveals an important tension: the very diversity within the visually impaired community can become a barrier to social inclusion, especially when shared references and sensory assumptions diverge. Such divergence limits the formation of mutual understanding and often leads to the creation of small, insular sub-communities based on impairment type, further fragmenting the broader community.

This stratification extends beyond conversation and identity—it affects access to information, strategies, and tools within mobile game communities. For example, V9, who became blind later in life, was able to draw on prior gaming experience, knowledge of mainstream platforms, and pre-existing networks to quickly re-integrate into various gaming communities, even those without full accessibility features. In contrast, most congenitally blind participants relied heavily on fully accessible games developed specifically for blind users, which significantly limited their exposure to broader player networks and mainstream gaming discourse.

These structural gaps often translated into unequal access to informal knowledge, such as assistive tools, in-game tactics, or community-based workarounds, that are often passed along through peer-to-peer channels. Participants who were excluded from mixed-impairment or mainstream groups reported being unaware of commonly used third-party tools or alternative game genres. Thus, internal segmentation not only fragments social bonds but also amplifies information inequality within an already marginalized population.

\subsubsection{The Challenges of Intersectional Inequality in Mobile Game}\label{5.3.2}

Beyond physiological and sensory differences, our findings reveal that intersectional identities, particularly those involving gender, nationality, and disability, compound the social marginalization faced by visually impaired players in mobile games. These intersecting systems of oppression reflect not only real-world social hierarchies, but also expose how digital gaming environments reproduce and sometimes intensify these inequalities.

Participants described how national identity became a source of stigma in global gaming spaces. V22 recounted her experience in international chat channels within blind-accessible games: \textit{Visually Impaired players from other countries often complain in global chats about being scammed after purchasing Chinese accounts. When they make complaints, few people oppose their complaints and seems admit Chinese are all bad, which may be due to the widespread prejudice against the Chinese people. This prejudice also extends to discrimination against our Chinese visually impaired mobile players."}

In such spaces, national stereotypes intersect with disability, leading to a form of layered exclusion, where players are not only othered because of their blindness, but also dismissed or distrusted due to their cultural identity. These biases are rarely challenged, further normalizing discriminatory discourse within so-called “accessible” communities.

Gender further complicates this picture. Among the 30 participants, 28 observed that female visually impaired players face heightened levels of harassment, both because of gender and disability. These players often experience misogynistic microaggressions, capability-based assumptions, and unsolicited attention, which collectively diminish their presence and visibility within the game. V17 shared: \textit{"Whenever I speak, people often say that a woman cannot be good at games. Sometimes, when I accidentally reveal that I am visually impaired, they look down on us even more."} 

This double marginalization silences players and suppresses their willingness to participate in open conversations. The result is not just discomfort but active withdrawal from digital communities. For example, V5 and V27 reported leaving game communities entirely to avoid repeated verbal harassment and gendered policing.

These experiences illustrate that “accessible” gaming environments are not inherently inclusive, especially when platform design fails to address the intersectional structures of power and vulnerability.

\section{Findings: Visually Impaired Players' Expectations for Socializing in Games (RQ3)}

In this section, we present the expectations of visually impaired players for more inclusive and socially meaningful mobile gaming experiences. While current games offer important avenues for communication and companionship, participants envisioned a future in which mobile games serve not only as accessible entertainment platforms but also as bridges to broader social connection. Their aspirations centered around three key areas: (1) inclusive design that enables shared play between different users, rather than segregated versions (\autoref{shared}); (2) interest-based, flexible social mechanisms that reduce both intra-group and inter-group barriers (\autoref{companion}); and (3) regionally informed systems that support safe and gradual transitions from online interaction to offline companionship (\autoref{match}). 

\subsection{Shared Gaming Spaces Between Different Players: Avoiding Version Segregation in Game Designing}\label{shared}

A recurring theme across our interviews was the strong desire for inclusive integration rather than separation. Nineteen visually impaired participants emphasized that when games are designed with accessibility in mind from the outset, there is no inherent reason for gameplay segregation. As V11 put it, despite physiological differences, skill disparities between sighted and visually impaired players can be minimal, what matters is whether the game environment supports equitable access. V14 echoed this view, stressing that mainstream game developers should include visually impaired users in their design assumptions, rather than relegating them to niche products or afterthoughts.

Several participants (n=10) shared experiences of playing games that offered separate versions, one for sighted users and one for blind or low-vision players. While intended as a form of inclusion, these parallel designs often result in social separation. V29 articulated this tension vividly: \textit{"Developing separate versions for visually impaired and sighted players is like having a dam between reservoirs that are otherwise connected. While living in the same environment, the division prevents integration. Games should break down this barrier, allowing everyone to play together and enabling socializing between us and sighted individuals. In fact, achieving this can be quite simple. Game developers only need to consider visually impaired players during the initial design phase, ensuring that the game is compatible with screen readers. Giving us the opportunity to try such games would already be a great step forward."}

For these participants, inclusion does not mean the creation of alternative experiences but means equitable access to shared worlds. Their vision is not of games that accommodate disability as an exception, but of games that recognize sensory diversity as part of the norm. This call from our participants challenges the logic of versional segregation and suggests that inclusive design should begin at the point of mainstream development, not as a post-hoc fix.

Moreover, participants also highlighted the need to recognize intra-group diversity among individuals with visual impairments. While often grouped together under a single accessibility label, visually impaired players vary widely in their needs, ranging from complete blindness to partial sight, and in their preferred modes of interaction, such as audio cues or haptic feedback. This diversity underscores the importance of flexible and customizable design features, rather than one-size-fits-all solutions. They expected truly inclusive gaming environments must go beyond simply accommodating “the blind” as a monolithic category and instead embrace the full spectrum of sensory differences within and beyond the visually impaired community.

\subsection{Inclusive Companion Models: Reducing Social Segregation Through Interest-Based Connection} \label{companion}

In-game socializing often remains confined to the gameplay context, with limited spillover into more sustained or diverse relationships. However, among our participants (9 visually impaired), the emerging popularity of "companion (\textit{Dazi})" models, a lightly-structured, interest-based form of social connection, was seen as a promising approach to fostering meaningful interaction across both intra-group and inter-group lines. Originating from recent popular culture among sighted youth in China, this socializing model describes relationships that form around shared activities or needs, often evolving into long-term companionship across domains like gaming, music, or studying ~\cite{luo2025impact}.

Several participants (n=5) expressed a strong desire for such practical and flexible social connections. They want to find a friend that could co- enjoy other areas from mobile gaming. V21 explained: \textit{"We don't go out much, so we usually make friends by chatting online. I enjoy playing games, so many of my friends are people I met through gaming. One close friend and I often discuss game mechanics and strategies, and when a new game is released, we team up to play together, becoming 'gaming companions.' I also hope to find a study companion from my special education school or a music companion with similar tastes through mobile gaming."}

This model offers emotional companionship and represents a low-pressure, high-affinity form of social interaction that is particularly well-suited to those with limited physical mobility or access to traditional social venues. Importantly, the companion model, according to our participants, offers several principles of inclusive social design: 1) Interest-first matching: By foregrounding shared interests rather than social identity or physical ability, the model minimizes the salience of disability as a social barrier. 2) Loose-tie, low-stakes interaction: Compared to fixed-role communities or identity-based groupings, companion platforms allow for scalable and modular connection based on interested areas, reducing fear of rejection or exclusion. 3) Cross-group permeability: Companionships can form between sighted and visually impaired individuals through co-experience (e.g., playing the same game), or between low-vision and blind individuals, helping to erode segregation and encourage genuine collaboration.

\subsection{Regional Matching: Expanding Offline Socializing} \label{match}

While mobile games provide a valuable channel for socializing, many participants expressed a deep longing for in-person connection, a type of social engagement that games alone cannot fully replace. Among the participants (n=11), several reported developing strong online friendships through gaming, with three attempting to meet their companions in person. However, only one—V1—successfully transitioned from online to offline social interaction, largely due to family support.

This difficulty points to a broader constraint: geographical distance often prevents the deepening of social bonds, particularly for visually impaired individuals who already face mobility and accessibility challenges. V26 emphasized this issue and proposed a region-based approach to matchmaking in games: \textit{"We have great online friends but rarely have the chance to meet. If we were closer, the likelihood of successful offline meetings would increase. Once successful cases emerge, the range of distance can gradually expand, increasing the chances of offline interaction for the visually impaired. If we can find companions, we would also be more willing to participate in social activities together."}

This model of regional socializing matching, if possible in our participants' expectations, enabling players to connect with others in the same city or region, could support the gradual, safe transition from online communication to offline companionship. Geographical proximity also provides functional benefits: it allows players to share information about local accessible facilities, public events, or disability-friendly social spaces. Several participants noted that they would feel more confident attending nearby events, such as eSports tournaments or casual game meetups, if they had companions they already knew and trusted through online play.

However, concerns about safety remained prominent. Eight participants expressed hesitation about attending offline events, especially those organized through informal channels. V1, who has experience with tabletop games and offline game groups for visually impaired people, recommended that formal institutions, such as schools, disability support centers or game platforms, take an active role in organizing and regulating offline gatherings. Participants (n=21) highlighted the need for clear safety protocols, including trained volunteer guides, accessible venues, and structured meeting arrangements. When these measures are in place, they reported feeling significantly more willing to participate in offline activities.

These findings reveal that for visually impaired players, games are not just a social space, they are even expected to be a bridge to re-entering public life for these visually impaired people.

\section{Discussion}

Our study reveals that for visually impaired individuals, mobile games are not only spaces for entertainment but also important platforms for building relationships, accessing information, and participating in social life. In this section, we reflect on the broader implications of our study, focusing on three key themes. First, we examine how mobile games can support more inclusive forms of social interaction and offer design insights to enhance accessibility and participation (\autoref{enhance}). Second, we discuss the role of structural ableism and social exclusion within gaming environments, highlighting the barriers that persist even in spaces labeled as “accessible” (\autoref{ableism}). Thirdly, we argue for a deeper recognition of mobile games as sites of social infrastructure, especially for marginalized communities who face limited access to other public and digital spheres (\autoref{mobile}). Finally, we outline several concrete design strategies based on our findings that aim to address relational inclusion in mobile gaming (\autoref{Design}).

\subsection{Enhancing Socializing through Mobile Games for Visually Impaired Individuals} \label{enhance}
Drawing from the first part of our findings, mobile games have emerged as a critical site for social participation among individuals with visual impairments. Beyond their entertainment value, mobile games function as alternative social infrastructures that help mitigate accessibility barriers and information asymmetries present in everyday life. For instance, in our study, in-game communication channels, such as real-time voice chat and team-based coordination, enabled visually impaired participants to access timely information that would otherwise be difficult to obtain through traditional means, due to physical or infrastructural constraints. This aligns with prior scholarship suggesting that digital platforms can act as compensatory spaces for marginalized groups, where online social engagement helps address the structural exclusions encountered in offline environments \cite{valkenburg2007online, kraut2002internet, zywica2008faces}.

Our findings extend this line of inquiry by highlighting the multifunctionality of mobile game–based social interaction. In addition to enabling emotional connection and peer bonding, these environments also serve as informal information networks, providing practical knowledge, everyday tips, and real-time updates among players. This reaffirms the role of mobile games not merely as leisure tools, but as socially generative systems that support broader forms of engagement and mutual care, particularly for individuals whose access to information and companionship is frequently constrained by systemic inaccessibility.

Moreover, our research illustrates how mobile games help cultivate a shared interactional space where sighted and visually impaired players can engage on more equitable terms. Drawing from Goffman’s theory of interactional framing ~\cite{persson2018framing} and the concept of “third places” proposed by Oldenburg \cite{oldenburg1999great}, mobile games can function as semi-anonymous, low-barrier social environments where the visibility of disability is minimized and identity can be reconstituted around gameplay roles, strategies, and shared goals. Third places refer to spaces outside of home (first place) and work (second place), such as cafes, parks, or clubs, where individuals gather for casual interaction, mutual support, and community building. These spaces are characterized by their inclusivity, voluntary participation, and the blurring of social hierarchies. In the context of mobile games, we observe similar qualities: players engage voluntarily, social roles are fluid, and shared goals often outweigh social distinctions such as disability status. Many participants described experiencing reduced self-consciousness and psychological stress during in-game socialization, in part because visual markers of difference, such as the use of assistive devices or altered mobility, were no longer salient or stigmatized in the virtual environment. This temporary suspension of real-world social hierarchies enabled more fluid interactions, moments of relational parity, and a stronger sense of belonging.

Our study contributes to a growing body of work in HCI and game studies that repositions games not only as sources of entertainment, but also as sites of social repair and relational agency. In doing so, we call for a deeper recognition of the inclusive potential embedded in everyday gaming practices and advocate for design approaches that actively center non-visual, collaborative, and emotionally supportive modes of interaction.

\subsection{Ableism in Mobile Game Socializing} \label{ableism}

Our findings reveal a complex paradox at the heart of mobile game–based socializing for visually impaired individuals: while these platforms can foster rich, meaningful, and even empowering forms of interaction, they also reproduce forms of marginalization rooted in ableist assumptions about who counts as a competent, desirable, or "normal" player.

Disability studies scholars have long emphasized the need for visibility justice, which is a condition in which marginalized individuals are not only seen but are recognized and engaged with on their own terms \cite{XNZS202201016}. Our participants, however, repeatedly described how gaining “visibility” in mobile games, whether through voice chat, performance, or social interaction, often risked being reabsorbed into dominant ableist frameworks. Visibility, in this sense, becomes double-edged: it offers access, but also exposure to judgment, misunderstanding, and symbolic violence.

The interviews in our study also surface layers of internalized ableism and intra-group stratification within the visually impaired community itself. This reflects broader critiques from critical disability studies and intersectionality theory: disability is never experienced in isolation, but is co-constituted by socioeconomic, technological, and cultural factors \cite{puar2017right}. These stratifications complicate any simplistic framing of the “visually impaired gamer” as a unified subject and push us to consider how mobile games are not only social spaces but also sites where disability itself is differentiated and negotiated.

In short, our study suggests that mobile games are not just entertainment platforms, since they are cultural and infrastructural spaces where disability, identity, and sociality are continually negotiated. Addressing ableism in these environments requires more than technical fixes: it calls for a reimagining of what counts as play, who is considered a player, and how social value is distributed across diverse bodies and ways of being.

\subsection{Why Mobile Games Matter: Rethinking Social Participation through Play} \label{mobile}

Our study highlights the importance of investigating mobile games not merely as tools of entertainment, but as emergent social infrastructures, especially for marginalized communities such as visually impaired individuals in Global South. Unlike traditional platforms for information access and social connection, which often privilege fast-paced, visual, and fleeting interactions, mobile games offer structured environments where communication unfolds more deliberately, with greater emphasis on voice, cooperation, and shared task orientation. These features create a relatively accessible and relationally rich space for social interaction.

Recent work by Ran et al. \cite{ran2025users} further underscores the global significance of mobile gaming for blind and low vision (BLV) communities, especially in the Global South. Their study highlights that, compared to desktop platforms or physical tabletop games, mobile devices are often more affordable, portable, and widely adopted in resource-constrained settings. In regions where infrastructural or economic barriers limit access to expensive assistive hardware or specialized gaming equipment, mobile phones serve as the primary—if not sole—portal to both gameplay and digital social life. This observation is particularly relevant in the Chinese context, where mobile-first platforms dominate online culture, and smartphones are often the most accessible digital device for disabled users ~\cite{he2023have,ran2025users}. Against this backdrop, examining the social dynamics of mobile gaming provides not only a culturally situated perspective, but also an infrastructurally grounded one that reflects the actual conditions through which visually impaired users engage with digital leisure.

Importantly, mobile games offer a form of mediated social parity that is rarely found in other digital contexts. Within game environments, visual impairments can become temporarily less salient, not because they are erased, but because the mechanics of play emphasize collaboration, strategy, and mutual reliance over visual speed or appearance. As participants in our study noted, it was often through multiplayer games that they first felt able to participate in peer groups on more equal terms. This echoes broader arguments in HCI and disability studies that call for moving beyond assistive fixes and toward the design of shared infrastructures of inclusion ~\cite{abascal2005moving}.

While prior research on disability and technology has predominantly focused on accessibility in formal domains such as education \cite{mikropoulos2023digital}, employment \cite{das2021towards}, and navigation \cite{carmien2005socio,zahabi2023design}, the domain of leisure, particularly mobile play, has received comparatively less sustained attention. Although a growing number of studies have begun to examine how disabled individuals engage with games and playful technologies, much of this work remains limited in scope or concentrated in Global North contexts. As a result, the everyday gaming practices of marginalized users, especially in underrepresented regions, continue to be under-studied and under-theorized. Yet, as our participants revealed, it is precisely through casual spaces like games that cultural knowledge circulates: jokes, news, slang, emotional support, and everyday chatter. For visually impaired players, mobile games thus serve as cultural contact zones, enabling participation in forms of sociality that are otherwise difficult to access due to platform inaccessibility or informational lag.

This line of inquiry also expands the conversation on social accessibility in HCI. Most accessibility work has focused on interface-level access or task completion \cite{maidenbaum2013increasing,collins2023guide,li2017braillesketch}, often overlooking the social dynamics of being included. By studying mobile game–based socialization, we foreground these affective and relational dimensions of access. We argue that inclusion is not simply a technical issue but a socio-technical achievement that is co-constructed through design, participation, and community norms.

\subsection{Design Implications: Towards Relational Inclusion in Future Mobile Game Design} \label{Design}

Our findings highlight that inclusive mobile game design must go beyond the narrow definition of accessibility as mere entry into a game. Instead, it should support what we term relational inclusion—the ability not only to participate, but to be heard, to feel a sense of belonging, and to co-create shared social experiences. This reframing aligns with broader calls in disability studies and accessible computing to prioritize the social and structural dimensions of inclusion, rather than focusing exclusively on technical fixes ~\cite{ellcessor2017disability,hamraie2017designing}.

Visually impaired participants often described mobile game communities as valuable spaces for connection, yet expressed frustration with exclusionary patterns that mirrored their broader social experiences. These insights suggest that inclusive game design must grapple with not just individual interaction challenges but also the systemic inequities that shape who gets to connect and thrive in game-based social spaces.

To that end, we propose several concrete design strategies based on our findings that aim to address relational inclusion in mobile gaming:

\subsubsection{Interest-Based Social Matching:} Matchmaking systems in mobile gaming can be redesigned to pair players, including both visually impaired and sighted individuals, as well as players with varying degrees of visual impairment, based on shared gameplay interests, complementary play styles, or social preferences. While accessible design often frames inclusion in binary terms (e.g., blind vs. sighted), our findings show that differences within the visually impaired community, such as between those congenitally blind individuals and those with residual vision \cite{ran2025users}, can also create social divides and feelings of disconnection. Interest-based matching helps mitigate these divides by foregrounding common affinities rather than perceived differences. By enabling players to connect through shared narratives, goals, or aesthetic preferences, this approach shifts the emphasis from functional ability to personal identity, allowing users to enter relationships on equal footing. Designing for cross-ability inclusion means facilitating opportunities for people to meet not as blind or sighted players, but as fellow gamers drawn together by mutual curiosity, enjoyment, and belonging.

\subsubsection{Region-Aware Companion Systems:} Companion systems embedded within mobile games can be designed to be regionally adaptive, recommending not only in-game events that are relevant to a player’s local context, but also accessible offline meetups such as community gatherings, low-vision-friendly game cafés, or audio-described public events. For many visually impaired players, particularly those in developing countries or rural areas, mobility constraints, poor infrastructure, and limited social support hinder participation in mainstream social life. These challenges are compounded by broader structural issues such as the lack of public transportation, social stigma, or inadequate orientation and mobility training. In our study, several participants emphasized that one of their primary motivations for joining mobile gaming communities was to make friends, the real, lasting friendships that go beyond the game itself. For these players, mobile games are not just a form of entertainment, but a rare entry point into social worlds from which they are often excluded. Region-aware companion systems can bridge virtual and physical worlds, enabling players to transition from online interactions to offline relationships in safe, supported, and localized ways. By surfacing geographically relevant events and connecting players who live near one another, these systems can help reduce the sense of isolation and disconnection that often accompanies disability in under-resourced settings. Thus, they reaffirm the role of mobile games as a meaningful infrastructure for social inclusion, not just in the digital realm, but in everyday life.

\subsubsection{Role-Diverse Team Structures: }Inclusive mobile games can introduce role-based mechanics that recognize and leverage players’ diverse strengths, such as strategic planning, navigation, storytelling, or social coordination. These roles can reflect individual interaction preferences (e.g., voice versus text communication), sensory modalities, or interpersonal styles (e.g., initiators, supporters, problem-solvers). For visually impaired players, being assigned a distinct and valued role within a team is not just a gameplay function but can also serve as an affirmation of competence and social worth. In our study, several participants expressed concern about being perceived as burdens member in mobile gaming. Such perceptions often lead to a tendency to “stay quiet” in group interactions. Role-diverse structures help counter this by building systems of interdependence: every player has a purpose, and success depends on the contribution of each member. This can reshape the social dynamics of gaming communities, transforming visual impaired players from passive observers to indispensable collaborators. When players are recognized for their unique strengths, whether it’s remembering map layouts, rallying team morale, or managing inventories, they are more likely to feel respected and included. These roles can open up new opportunities for connection and friendship. Social bonds often form more readily when players feel needed and appreciated within a group. By giving all players a chance to be seen as skilled, supportive, role-diverse team structures can foster stronger peer-to-peer relationships and enhance self-esteem. 

These three strategies are informed by the lived experiences and expectations of our participants, shifting the focus from merely enabling access to fostering equity and meaningful engagement. While existing accessible game design has largely focused on technical innovations, such as tactile and auditory feedback ~\cite{allman2009rock,mangiron2016game,qiu2025gamerastra}, our work takes the expectations and lived experiences of visually impaired participants as a starting point to explore how specific game mechanisms can be designed to support the social and infrastructural dimensions of inclusion. We hope these implications provide practical guidance for designers and contribute conceptually to disability and mobile gaming research, reframing inclusive design not as the accommodation of impairment, but as the redesign of play’s social architecture.

\section{Limitations and Future Work}
Despite providing valuable insights into the social needs and gaming experiences of visually impaired players, our research has several limitations. First, all participants in our study were based in China, and cultural, infrastructural, and technological differences may shape accessibility perceptions and practices in other regions. This may limit the generalizability of our findings to the global population of over 2.2 billion individuals with visual impairments ~\cite{rizzo2023global}. 

Second, this study focuses primarily on the experiences and perspectives of visually impaired players themselves. While this lens foregrounds marginalized voices, future work would benefit from a broader ecological perspective. Incorporating the views of other stakeholders, such as sighted players and game developers, can offer a more holistic understanding of how inclusion and exclusion are co-constructed in multiplayer and cross-platform environments. Such perspectives would be particularly valuable in understanding interpersonal dynamics, design trade-offs, and the cultural norms that shape who gets to participate and how. 

Third, future studies could explore longitudinal or intervention-based methods to assess how changes in game design, such as the introduction of inclusive features, affect players' experiences and social integration over time. 

Finally, we see important opportunities for expanding this line of research beyond visual impairments. Future work could investigate how game accessibility intersects with other dimensions of disability and marginalization, such as auditory or cognitive impairments, older adults with changing sensory needs, or players in resource-constrained environments. Researchers and designers can work toward building truly inclusive gaming spaces that recognize and adapt to the full spectrum of human difference ~\cite{qiu2025gamerastra}.

\section{Conclusion}
Through semi-structured interviews with 30 visually impaired players, this study explores the social experiences of visually impaired individuals in mobile gaming. We distinguish how visually impaired players engage in socialization through mobile games, including in-game social systems, guilds, and external relationships and social activities. While mobile games serve as a crucial avenue for social interaction among visually impaired individuals, significant barriers remain, including technological limitations, challenges in gameplay, and reflections on real-world constraints within virtual environments. Additionally, we analyze visually impaired players' expectations for social interaction design in future mobile games. We hope future researchers will continue to explore the social experiences of various marginalized groups in mobile gaming and examine the dual role of mobile games in both social enhancement and compensation. 

\bibliographystyle{ACM-Reference-Format}
\bibliography{reference.bib}

\appendix
\end{document}